# Effect of Electron-Phonon Interaction and Ionized Impurity Scattering on the Room Temperature Thermoelectric Properties of Bulk MoSe$_2$

Md Golam Rosul[1], Mona Zebarjadi[1,2,*]

[1] Department of Electrical and Computer Engineering, University of Virginia, Charlottesville, VA 22904, United States

[2] Department of Materials Science and Engineering, University of Virginia, Charlottesville, VA 22904, United States

[*]Corresponding author: m.zebarjadi@virginia.edu

**Abstract**

We study the thermoelectric properties of bulk MoSe$_2$ within relaxation time approximation including electron-phonon and ionized impurity interactions using first-principles calculations at room temperatures. The anisotropy of this two-dimensional layered metal dichalcogenide is studied by calculations of electron mobility in the cross-plane and the in-plane directions. We show that the cross-plane mobility is two orders of magnitude smaller than the in-plane one. The inclusion of van der Waals interactions further lowers the carrier mobility in the cross-plane direction but minimally affects the in-plane one. The results for in-plane electrical mobility and conductivity are in close agreement with experimentally reported values indicating the accuracy of the calculations. The Seebeck coefficient calculations show that this coefficient is primarily dictated by the band structure. The details of relaxation times and inclusion of van der Waals interactions only slightly change the Seebeck coefficient. The in-plane thermoelectric power factor reaches a maximum value of 20 µWcm$^{-1}$K$^{-2}$ at a carrier concentration of 1.5x10$^{20}$ cm$^{-3}$ at 300K.

**Introduction**

Thermoelectric materials can convert heat energy directly to electricity and vice versa. Thermoelectric materials are considered to have great potential for power generation, energy-saving, and heat management.[1-6] Devices made of thermoelectric materials are extremely simple, have no moving parts, produce no greenhouse gases, operate quietly, are environmentally friendly, and are highly reliable.[7,8] Because of these excellent features, thermoelectric devices have attracted extensive interest for nearly two centuries. Solid-state thermoelectric devices are generally based on heavily doped semiconductors such as chalcogenides[9,10], zintl phases[11], clathrates[12], complex oxides[13], and skutterudites[14,15], and can be used for cooling applications or electricity generation directly from a heat source. The efficiency of these devices is determined by a dimensionless figure-of-merit $ZT = (S^2\sigma/k)T$ where $Z$ is the figure-of-merit, $T$ is the absolute temperature, $S$ is the Seebeck coefficient, $\sigma$ is the electrical conductivity, and $k$ is the total thermal conductivity with contributions from the lattice ($k_L$) and the electrons ($k_e$).[16] Besides the traditional power generator and cooling applications, thermoelectric materials can also be applied to active cooling[17,18], where large thermal conductivity and power factor are both desired to transfer heat from high-temperature heat sources to low-temperature heat sinks, and to thermal switches, where heat flux is required to adjust based on ambient conditions[19].

Two-dimensional (2D) layered materials such as transition metal dichalcogenides (TMDCs) are good candidates for thermoelectric [20–28] as well as thermionic[29–32] energy conversion applications because of their large Seebeck coefficients and low thermal conductivities. Due to weak van der Waals interactions, the thermal conductivity in the cross-plane direction is small making them ideal candidates for nanoscale cooling[33–35] and power generation. Another important feature of 2D materials is that the thermoelectric properties such as electrical conductivity and Seebeck coefficient depend on the number of layers as the band structure and the bandgap change as the thickness of the material varies. In the case of $MoSe_2$, its bulk has an indirect bandgap of 0.80 eV while monolayer $MoSe_2$ has a direct bandgap of 1.55 eV.[36–38]

$MoSe_2$, a TMDC material, consists of a transition metal Mo, sandwiched between two chalcogen layers of Se in which Mo's and Se's are covalently bonded within the plane. However, the Se layer constructs weak van der Waals (vdW) interaction with the next Se layer perpendicular to the plane to construct bulk $MoSe_2$. Therefore, it is important to include the van der Waals interaction in the theoretical calculation to correctly obtain the thermoelectric transport properties of $MoSe_2$. It is expected that the inclusion of vdW interaction in the theoretical calculation affects the cross-plane transport properties significantly.

Thermoelectricity in semiconductors is the response of electron and phonon currents to temperature gradients. The interaction between the electrons and phonons plays a crucial role in this response. To maximize the thermoelectric response, one needs to selectively heat electrons and minimize the electron-phonon interaction to avoid heat leakage to the lattice. Only the energy carried by the electrons contributes to the conversion of heat to electrical energy, the part carried by phonons is wasted. In practice, phonons always exist at finite temperatures and take some of the input heat directly from the source and some through electron-phonon energy exchange. In either case, phonon contributes to heat leakage that lower the performance of the material. Electron-phonon interaction is an important phenomenon in condensed matter physics beyond thermoelectricity. Many experimental observations such as temperature-dependent band structures, zero-point renormalization of the bandgap in semiconductors, conventional phonon-mediated superconductivity, phonon-assisted light absorption, Peierls instability[39], the Kohn effect[40], temperature-dependent electrical resistivity as well as traditional superconductivity[41] are caused by the electron-phonon interaction. The role of electron-phonon interactions in the transport properties of systems with strong electron-phonon correlations is one of the central issues in the theory of strongly correlated systems.

Electron-phonon scattering plays a central role in electron transport in relatively pure materials.[42] While impurities and defects lower electron mobility, their values are not intrinsic and depend on the quality of the material growth and the number of impurity and defect centers. In contrast, electron-phonon interactions are an intrinsic property of a given semiconductor and hence the first step in evaluating the potential of a semiconductor for thermoelectric applications is to evaluate the electron-phonon interaction. Over the years, several different open-source codes have been developed to compute the electron-phonon scattering rate from first-principles calculations.[43–47] These fully first-principles approaches to calculating the electron-phonon interaction employing density functional perturbation theory (DFPT) combined with Wannier interpolation[48] can now produce highly accurate electron lifetimes and have shown close agreement with experimental measurements of electron mobility and conductivity of intrinsic samples.[44,46]

In the case of doped semiconductors, it is important to include ionized impurity scattering in addition to electron-phonon interactions. The carrier density in a doped semiconductor is determined by the relative position of the impurity states to the band edges (the binding energy) and the temperature. When ionized, the charged impurity centers scatter the free carriers via Coulomb interactions, affecting their lifetimes. At low temperatures, when phonon effects are minimal, carrier scattering by ionized impurities is the dominant scattering determining the carrier mobility. In heavily doped semiconductors, this regime might be extended to room temperatures. Therefore, it is important to incorporate ionized impurity scattering along with electron-phonon scattering in the theoretical transport calculation to accurately predict experimental results.

In this work, we theoretically evaluate the effect of electron-phonon scattering (EPS) and ionized impurity scattering (IIS) on the electronic and thermoelectric properties of bulk $MoSe_2$ in the in-plane and cross-plane direction at room temperature without and with including the effect of vdW interaction. We obtain the EPS rates from the first-principle calculations and the IIS rates using Brooks-Herring[49] approach. The electronic and thermoelectric transport properties such as mobility ($\mu$), electrical conductivity ($\sigma$), electronic thermal conductivity ($\kappa_e$), and Seebeck coefficient *(S)* of bulk $MoSe_2$ at room temperature in the in-plane and cross-plane direction are calculated by solving the linearized electron Boltzmann transport equation (BTE) under the relaxation time approximation (RTA) with the aid of the EPS rates and IIS rates evaluated at different Fermi Level positions. Then we compare the calculated in-plane mobility and electrical conductivity with experimentally obtained values. Finally, we use our developed method to optimize the in-plane power factor times temperature (PFT) of $MoSe_2$ with respect to carrier concentration and temperature.

**Computational Details**

We compute the EPS rates in bulk $MoSe_2$ using first-principles calculations. The equilibrium properties of electrons and phonons are calculated using density functional theory (DFT) and density functional perturbation theory (DFPT) as implemented in the QUANTUM ESPRESSO package.[50] The norm-conserving pseudopotentials[51] with the Perdew-Burke-Ernzerhof (PBE)[52] functional for the exchange-correlation are used. We used the non-local van der Waals DFT functional (vdW-DF-optb86)[53,54] to correctly take the vdW interaction into account. A 12×12×4 mesh and an 18×18×6 Monkhorst-Pack k-point mesh are used for the self-consistent and non-self-consistent field calculations, respectively and the cutoff energy of the plane wave is chosen as 60 Ry. The convergence threshold of energy is set to be $10^{-12}$ Ry. The lattice was relaxed with the force convergence threshold of $1.0e^{-4}$ Ry/Bohr. The obtained relaxed lattice constants of bulk $MoSe_2$ in the hexagonal structure are a=b=3.31 Å and c=12.89 Å. Our calculated lattice constants match the experimentally reported value in the literature.[55] The dynamical matrices and phonon perturbations are computed on a 6×6×2 q-point mesh in the phonon calculations. To obtain the EPS rates, the PERTURBO package[46] is employed to interpolate the electron-phonon coupling matrices as well as electron and phonon eigenvalues obtained by DFT and DFPT calculations from coarse to fine k and q point meshes (40×40×40) using the Wannier interpolation scheme.[56] The Wannier function consists of $d_{xy}$, $d_{xz}$, $d_{yz}$, $d_{z2}$, $d_{x2-y2}$ orbitals on each Mo atom and $p_x$, $p_y$, $p_z$ orbitals on each Se atom, leading to the wannierization of a total of 22 bands. In addition, and for comparison and validation, the EPS rates are also computed using the electron-phonon Wannier (EPW)[44] package based on maximally localized Wannier functions[56,57], which allows accurate interpolation of electron-phonon couplings from coarse grids to arbitrarily dense grids. The EPS rates calculated using these two packages show similar values. We calculated the IIS rates using the Brooks-Herring approach for different doping concentrations.[49] The IIS rate is given by,

$$\tau_{imp}(E) = \frac{8\pi^3\hbar^3\varepsilon_r^2\varepsilon_0^2}{N_I q^4}\left[\ln(1+Y^2) - \frac{Y^2}{1+Y^2}\right]^{-1} DOS(E)v^2(E)$$

where $Y^2 = 4\pi^2 L_D^2 \hbar DOS(E)v(E)$, $\hbar$ is the reduced Plank constant, $\varepsilon_r$ is the relative permittivity, $\varepsilon_0$ is the permittivity of free space, $N_I$ is the impurity concentration, $q$ is the electronic charge, $L_D$ is the Debye length, $DOS(E)$ is the density of states, and $v(E)$ is the group velocity.

The electron mobilities, electrical conductivities, electronic thermal conductivities, and the Seebeck coefficients are calculated using the PERTURBO[46] code based on the BTE under the RTA.

**Result and Discussion**

The electronic band structures of bulk $MoSe_2$ with and without the inclusion of vdW interaction are shown in Fig. 1(a). The band structure calculated in this work is in agreement with the previous calculation[58]. Bulk $MoSe_2$ shows an indirect bandgap of 0.80 eV. This is in agreement with the previously reported theoretical value.[37,38] The calculated phonon dispersions using DFPT with and without the inclusion of vdW interaction are shown in Fig. 1(b). The DFT band structure was then interpolated using the maximally localized Wannier function. The DFT calculated band structures and the band structure obtained from Wannier interpolation show very good agreement (see Supplemental Information). We observe that the effect of vdW interactions is minimal on the band structure and the phonon dispersions.

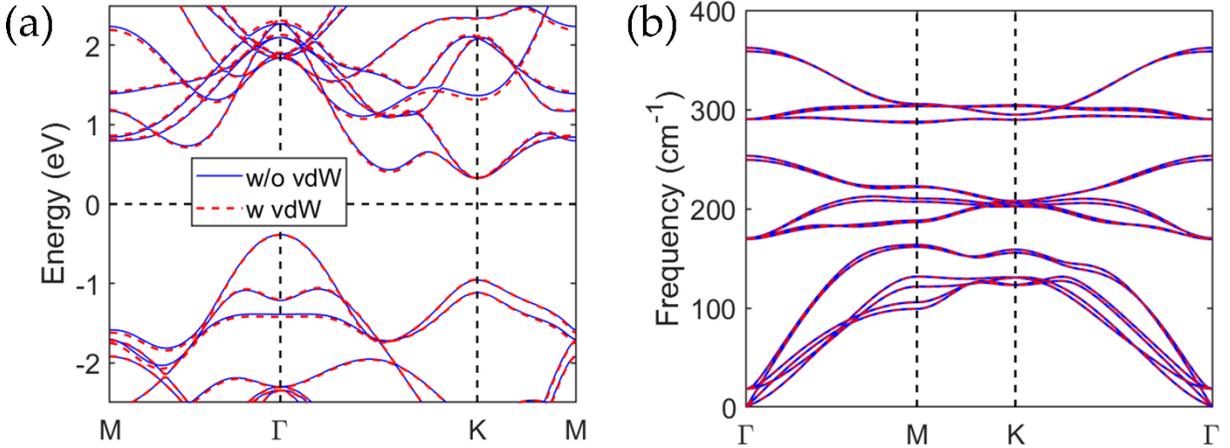

Figure 1. (a) Electronic band structure and (b) phonon dispersion of bulk MoSe$_2$ without (blue solid line) and with (red dashed line) the inclusion of vdW interaction.

We focus on the n-type samples and study the electron-phonon interaction within the conduction band. Figure 2 (a) shows the calculated EPS rates obtained from the EPW and PERTURBO codes together with the electronic density of states (DOS) with respect to the conduction band minimum (CBM). The EPS rates obtained from these two codes show close agreement. The scattering rates versus electron energy show a similar trend to that of the electronic DOS, consistent with the fact that the DOS regulates the phase space for electron-phonon scattering[59]. The IIS rates for the carrier concentration of $10^{16}$, $10^{17}$, and $10^{18}$ cm$^{-3}$ at 300K along with the EPS rates are shown in Fig. 2 (b).

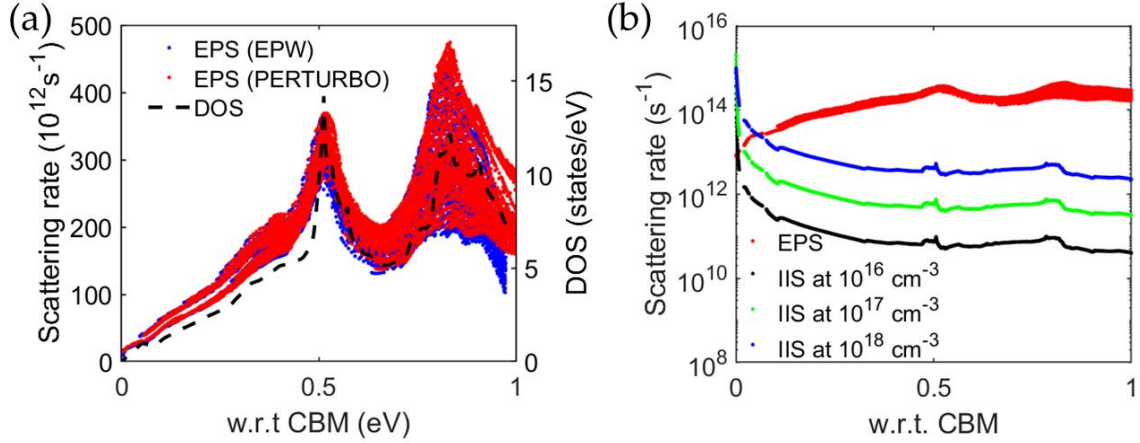

Figure 2. (a) Electron-phonon scattering rates in MoSe$_2$ calculated using EPW and PERTURBO codes together with the electronic density of states. (b) Electron-phonon scattering rates and ionized-impurity scattering rates with carrier concentrations between $10^{16}$ and $10^{18}$ cm$^{-3}$ at 300 K.

Using our calculated scattering rates, the room-temperature electron mobilities, electrical conductivities and electronic thermal conductivities are calculated using the BTE under the RTA. In this paper, we mainly focus on electron transport in the in-plane and cross-plane directions at room temperature. The electron transport properties, namely the mobility, electrical conductivity, and electronic thermal conductivity, are calculated for four combinations as shown in Figs. 3 and 4: (i) without vdW interaction and inclusion of only EPS rates (red curves), (ii) with vdW interaction and inclusion of only EPS rates (blue curves), (iii) without vdW interaction and inclusion of the combination of EPS and IIS rates (green curves), and (iv) with vdW interaction and inclusion of the combination of EPS and IIS rates (black curves). Figure 3 shows the room temperature electron mobility in the in-plane and cross-plane directions for the carrier concentration range of $10^{16}$ to $10^{18}$ cm$^{-3}$. Our calculated in-plane electron mobility at room temperature with the inclusion of only EPS rates and without vdW interaction is 192 cm$^2$/Vs while including the vdW interaction is 200 cm$^2$/Vs. These values are independent of the carrier concentration. The obtained cross-plane mobility with the inclusion of only EPS rates and without including the vdW interaction is 2.47 cm$^2$/Vs while with including the vdW interaction is 1.46 cm$^2$/Vs. We can see that the effect of vdW interaction on the in-plane mobility is small while the cross-plane mobility decreased by a factor of 1.7. This is expected as Mo's and Se's are covalently bonded within the plane while the Se layer constructs weak vdW interaction with the next Se layer in the cross-plane direction in the MoSe$_2$ structure. Thus to accurately predict the electronic transport in the cross-plane direction this vdW interaction needs to be accounted for in the calculations.

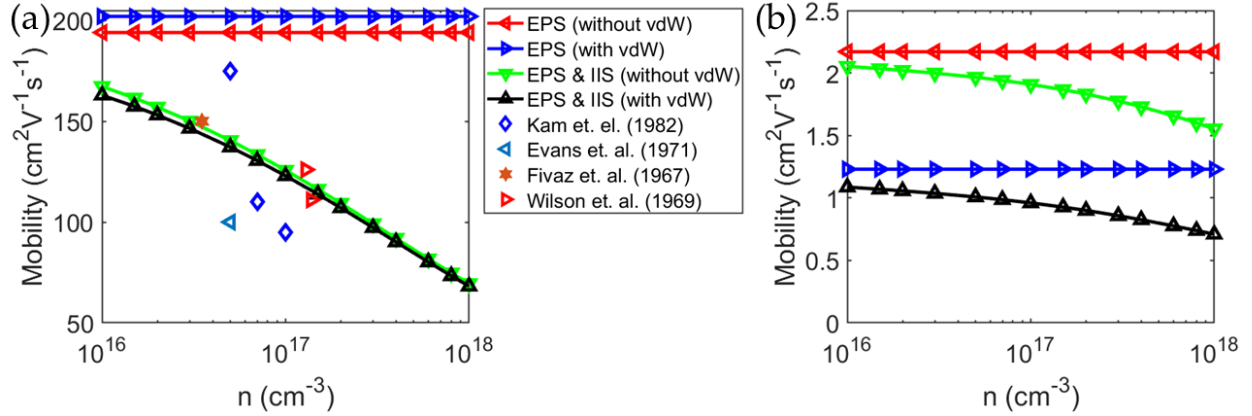

Figure 3. The (a) in-plane and (b) cross-plane mobility vs carrier concentrations are calculated using the Boltzmann transport equation under the constant relaxation time approximation. The in-plane electrical mobilities are compared with the in-plane experimental Hall mobilities from literature.

The theoretical mobility with the inclusion of only EPS rate is relatively insensitive to carrier concentration within the carrier concentration range of 10$^{16}$ to 10$^{18}$ cm$^{-3}$. The experimentally measured in-plane room temperature Hall mobilities of single-crystal MoSe$_2$ are extracted from literature for different carrier concentrations. Within the range of carrier concentration from $3\times10^{16}$ to $1.4\times10^{17}$ cm$^{-3}$, the Hall mobility varies from ~55 to ~175 cm$^2$/Vs as shown in Fig 3(a). The calculated in-plane mobility is overestimated compared to all the experimentally measured values. This is because we only included the EPS rates in the transport calculation which is the dominant scattering mechanism at low carrier concentration. However, as the carrier concentration increases, the contribution of the IIS rates in the transport properties is no longer negligible. Therefore, we need to include both the EPS rates and IIS rates in the transport calculation to correctly predict the experimental results. We can see from Fig. 3 that with the inclusion of both EPS rates and IIS rates in the transport calculation, mobility gradually decreases as the carrier concentration increases in the in-plane and cross-plane direction. Furthermore, we can see from Fig. 3a that the in-plane theoretical mobility closely matches the experimental mobility with the inclusion of both EPS rates and IIS rates thus confirming that the IIS rates need to be accounted for in the transport calculation to accurately predict electronic transport properties.

Next, we calculate the in-plane and cross-plane electrical conductivities at room temperature for the same four combinations. The calculated in-plane and cross-plane room temperature electrical conductivity for carrier concentrations from $10^{16}$ to $10^{18}$ cm$^{-3}$ are shown in Fig. 4(a) and 4(b). First, we calculate the in-plane and cross-plane electrical conductivities with the inclusion of only EPS rates. The in-plane and cross-plane electrical conductivities increase as the carrier concentrations increase, which is consistent with the behavior of semiconductor materials. The inclusion of vdW interaction does not affect the in-plane electrical conductivity as shown by the overlap of red and blue lines in Fig. 4 (a). However, the vdW interaction reduces the cross-plane electrical conductivity as shown in Fig. 4 (b). We compare our calculated in-plane electrical conductivity with the experimentally measured in-plane electrical conductivity of single crystal MoSe$_2$.[60–63] Our calculated in-plane values are larger than the experimental values as shown in Fig. 4(a). This is consistent with the fact that other scattering mechanisms are present in real materials along with EPS whereas we only included the EPS rates in the transport calculation. Therefore, we included the combination of both EPS rate and IIS rate in the transport calculation. We can see from Fig. 4(a) and 4(b) that the inclusion of IIS rates along with EPS rates has a small effect on the electrical conductivity at low carrier concentration but significantly modifies it as the carrier concentration becomes large. Now, our calculated in-plane electrical conductivity values closely match the experimental values as shown in Fig. 4(a) implying that the inclusion of both the EPS rate and IIS rate is important in the transport calculation of bulk MoSe$_2$ at room temperature.

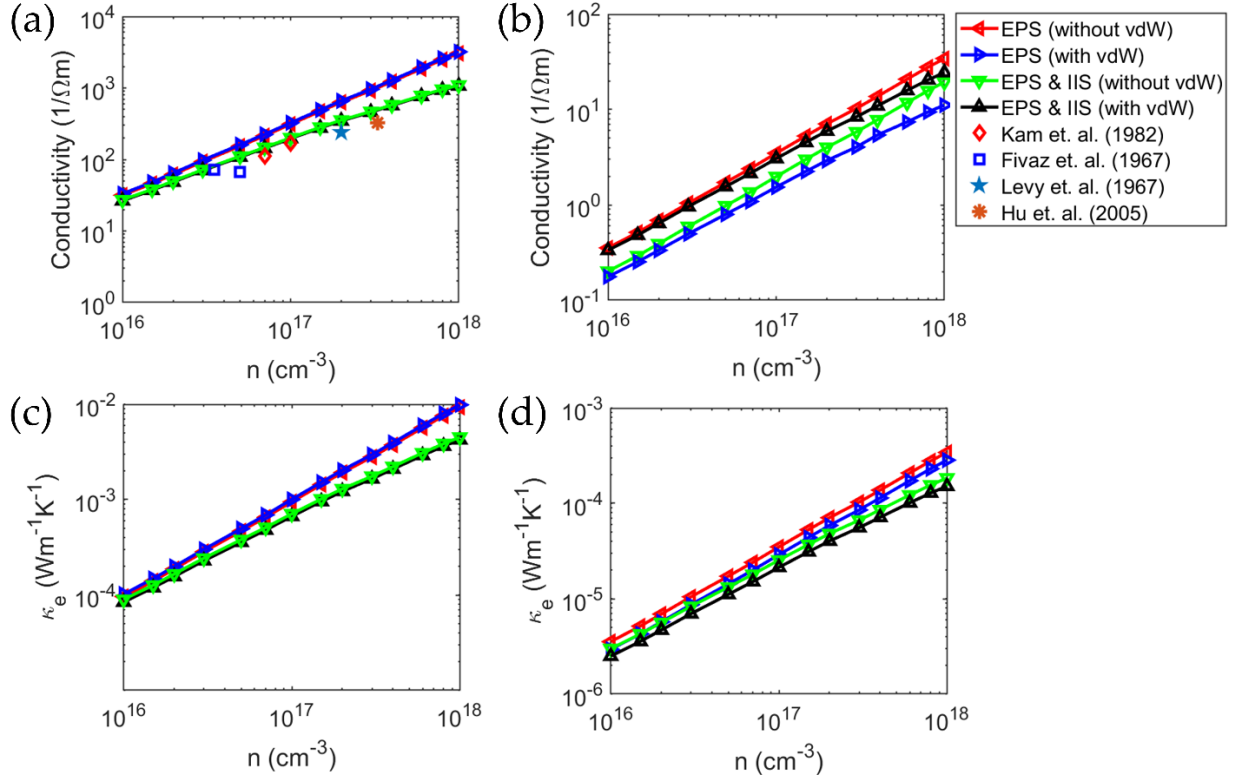

Figure 4. The (a) in-plane electrical conductivity, (b) cross-plane electrical conductivity, (c) in-plane electronic thermal conductivity (d) cross-plane electronic thermal conductivity vs carrier concentrations are calculated using the Boltzmann transport equation under the constant relaxation time approximation. The in-plane electrical conductivities are compared with the experimental electrical conductivities of single crystal MoSe$_2$ from the literature.

Then, for the same four combinations, we calculate the in-plane and cross-plane electronic thermal conductivities at room temperature. Figures 4(c) and 4(d) show the computed in-plane and cross-plane room temperature electronic thermal conductivity for carrier concentrations between $10^{16}$ and $10^{18}$ cm$^{-3}$. The electronic thermal conductivities are affected similarly to the electrical conductivities by vdW interaction, electron-phonon interaction, and ionized impurity scattering. This is consistent with the Wiedemann–Franz law. The lattice thermal conductivity is not calculated in this work. It is shown, however, that the lattice thermal conductivity is significantly reduced by the electron-phonon interaction at high doping concentrations for instance in silicon.[64]

The in-plane and cross-plane Seebeck coefficients are calculated using the BTE under the RTA. In the case of RTA, first, only the theoretically calculated EPS rates from first-principles are used to estimate the relaxation time. To carry out the quantitative comparison, we calculate the Seebeck coefficient using the constant relaxation time approximation (CRTA) employing the BoltzTrap code.[65] Figures 5(a) and 5(b) show the room temperature in-plane and cross-plane Seebeck coefficient vs chemical potential with respect to the conduction band minimum (CBM) calculated using RTA and CRTA. The Seebeck coefficients from RTA and CRTA show very good agreement confirming that the Seebeck coefficient is not sensitive to the details of relaxation times. The calculated Seebeck coefficients with and without including the vdW interaction are also shown in Fig. 5. The Seebeck coefficient value decreases as the chemical potential increases, which is consistent with the fact that higher chemical potential represents higher carrier concentration which contributes to the reduction of the Seebeck coefficient. The cross-plane Seebeck coefficient is larger than the in-plane Seebeck coefficient for the same chemical potential. In the case of the Seebeck coefficient, there is no effect of vdW interaction in the in-plane direction while there is a small but insignificant effect in the cross-plane direction. Agarwal et al. reported experimentally measured room temperature Seebeck coefficient of -960 µV/K at a carrier concentration of $1.78 \times 10^{16}$ cm$^{-3}$.[66] The chemical potential that corresponds to this carrier concentration is 0.21 eV below the CBM. In Fig. 5(a), the experimental Seebeck coefficient is shown as a red star. The calculated Seebeck coefficient of -1035 µV/K at room temperature at the chemical potential of 0.21 eV below the CBM is a reasonable estimate of the experimentally measured value. The inclusion of the combination of both the EPS rates and IIS rates in the Seebeck calculation does not have any effect on the Seebeck coefficients as shown by the black dashed line in Fig. 5 further confirming the insensitiveness of the Seebeck coefficient on the details of relaxation times.

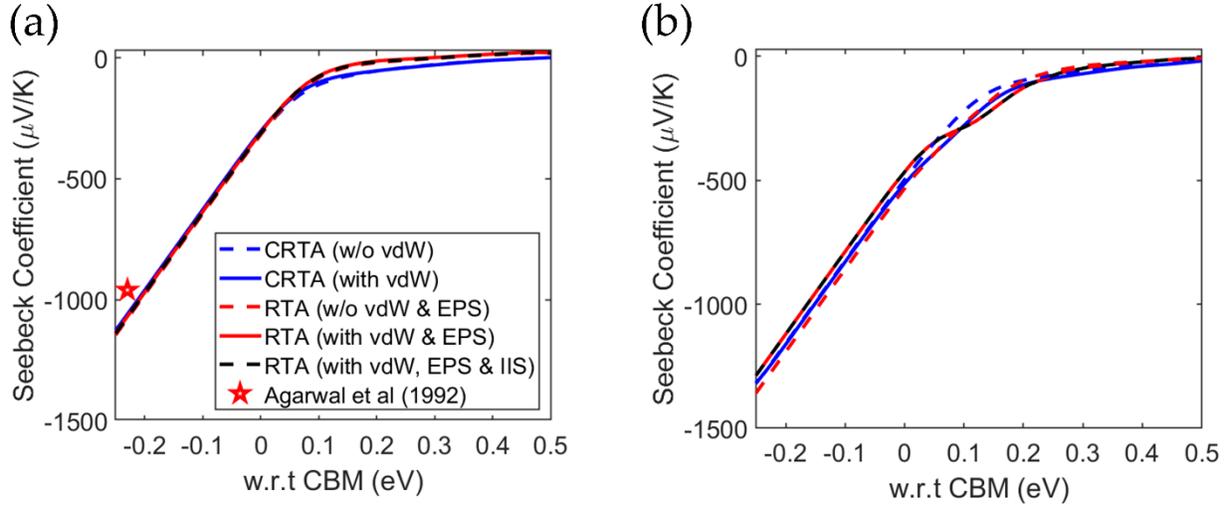

Figure 5. Room temperature Seebeck coefficient vs chemical potential in the (a) in-plane and (b) cross-plane directions. The blue (red) dashed and solid lines represent Seebeck coefficients calculated using CRTA (RTA including only the EPS rates) without and with vdW interaction respectively. The black dashed line represents Seebeck coefficients calculated using RTA with vdW interaction and the inclusion of the combination of EPS and IIS rates. The conduction band minima calculated from Boltztrap and Perturbo code are taken as the reference point, zero. The red star represents the experimentally measured Seebeck coefficient.

After reproducing the experimental in-plane transport properties of MoSe$_2$ with reasonable accuracy by including the vdW interaction in the structure relaxation and considering the combination of EPS and IIS rates in the transport calculation under RTA, we used our developed method to optimize the PFT with respect to carrier concentration and temperature. Figure 6(a) shows the optimized in-plane power factor with respect to carrier concentration at 300K. The in-plane PFT reaches a maximum value of 0.6 Wm$^{-1}$K$^{-}$ [1] at a carrier concentration of $1.5 \times 10^{20}$ cm$^{-3}$ at 300K. Then we fixed the carrier concentration at $1.5 \times 10^{20}$ cm$^{-3}$ and varied the temperature from 300K to 1200K. Figure 6(b) shows in-pane PFT vs temperature at a carrier concentration of $1.5 \times 10^{20}$ cm$^{-3}$. The PFT shows a maximum value of 1.07 Wm$^{-1}$K$^{-1}$ at 1000K.

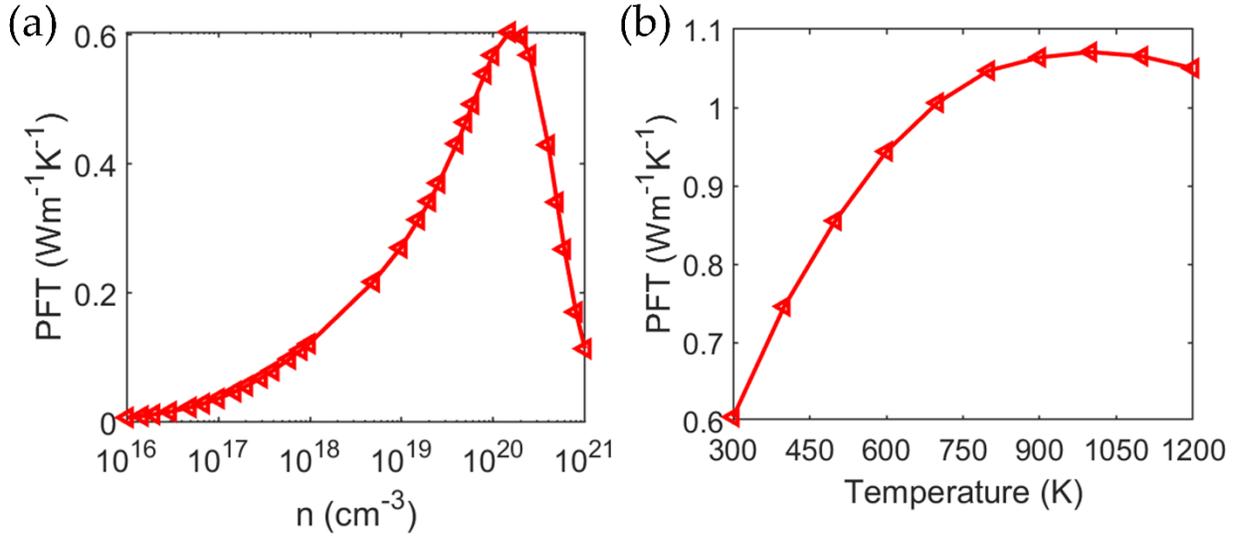

Figure 6. (a) In-plane power factor times temperature vs carrier concentration of bulk MoSe$_2$ at 300K. (b) In-plane power factor times temperature vs temperature of bulk MoSe$_2$ at a carrier concentration of $1.5 \times 10^{20}$ cm$^{-3}$.

**Conclusion**

In summary, we calculated the electron-phonon scattering rates of bulk MoSe$_2$ using two full first principle-based codes, PERTURBO and EPW. The ionized impurity scattering rates are calculated using the Brooks-Herring method. The calculated rates are then used to obtain the in-plane and the cross-plane electronic and thermoelectric properties of bulk MoSe$_2$ at room temperature with and without including the vdW interaction using the BTE under the RTA. The calculated in-plane and cross-plane electron mobilities with the inclusion of only electron-phonon scattering rates at room temperature without and with incorporating the vdW interaction are relatively insensitive to carrier concentration for a carrier concentration range of $10^{16}$ to $10^{18}$ cm$^{-3}$. The in-plane mobilities are not affected by the vdW interaction while the cross-plane mobilities are reduced by 1.7 times, thus confirming that the inclusion of vdW interaction is important to accurately predict the cross-plane transport properties of MoSe$_2$. The theoretically calculated in-plane electron mobilities and electrical conductivity values are overestimated compared to the experimental mobilities when electron-phonon scattering rates are only included. At high doping levels, electron transport is affected by ionized impurity scatterings in addition to the electron-phonon scattering. We were able to reproduce the experimentally measured in-plane electron mobilities and electrical conductivities after the inclusion of both rates and with no fitting parameters. The Seebeck coefficients are calculated by solving the linearized electron BTE under the RTA with the aid of the first-principles electron-phonon scattering rates evaluated at different Fermi level positions. To compare the calculated Seebeck coefficient under RTA, the Seebeck coefficients are again calculated using BTE under CRTA. The RTA and CRTA Seebeck coefficients show a good agreement indicating that the Seebeck coefficient is not sensitive to the details of the relaxation time. The inclusion of both the electron-phonon scattering rates and the ionized-impurity scattering rates in the Seebeck calculation did not influence the Seebeck coefficients, further proving the insensitivity of the Seebeck coefficient to the specifics of relaxation times. Finally, we used our developed method to optimize the PFT in terms of carrier concentration and temperature. We find that MoSe$_2$ shows a maximum PFT of 1.07 Wm$^{-1}$K$^{-1}$ at 1000K with a carrier concentration of 1.5x10$^{20}$ cm$^{-3}$.

**Conflicts of interest**

There are no conflicts to declare.

**Acknowledgments**

This work is supported by NSF grant number 1653268. The authors acknowledge the Rivanna High-Performance Computing system of the University of Virginia.

**Supporting Information**

Comparison of band structures obtained from DFT and Wannier interpolation.

# Supporting Information: Effect of Electron-Phonon Interaction and Ionized Impurity Scattering on the Room Temperature Thermoelectric Properties of Bulk MoSe$_2$

Md Golam Rosul[1], Mona Zebarjadi[1,2,*]

[1]Department of Electrical and Computer Engineering, University of Virginia, Charlottesville, VA 22904, United States

[2]Department of Materials Science and Engineering, University of Virginia, Charlottesville, VA 22904, United States

[*]Corresponding author: mz6g@virginia.edu

**Quantum Espresso and Wannier band structure:**

The maximally localized Wannier function was used to interpolate the DFT band structure. In the Wannier function, each Mo atom has $d_{xy}$, $d_{xz}$, $d_{yz}$, $d_{z^2}$, $d_{x^2-y^2}$ orbitals, whereas each Se atom has $p_x$, $p_y$, $p_z$ orbitals, resulting in the wannierization of a total of 22 bands. Figure S1 shows the band structure calculated from DFT and by interpolation using the maximally localized Wannier function. Both of these band structures show good agreement.

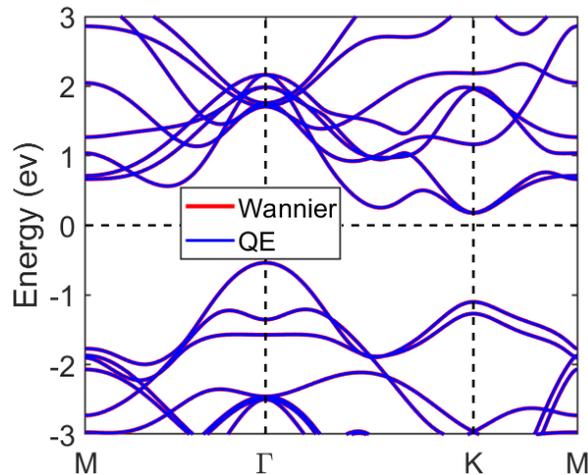

Figure S7. The band structure of MoSe$_2$. Results from DFT calculations (blue solid lines) and Wannier interpolation (red solid lines) are overlaid.